\documentclass[preprintnumbers,amsmath,amssymbm,preprint]{revtex4}
\usepackage{epsfig}
\usepackage{graphicx}

\begin{document}
\title{Analytic study of the Maxwell electromagnetic invariant in spinning and charged 
Kerr-Newman black-hole spacetimes}
\author{Shahar Hod}
\affiliation{The Ruppin Academic Center, Emeq Hefer 40250, Israel}
\affiliation{ }
\affiliation{The Hadassah Institute, Jerusalem 91010, Israel}
\date{\today}

\begin{abstract}
\ \ \ The Maxwell invariant plays a fundamental role in the mathematical description of electromagnetic fields 
in charged spacetimes. 
In particular, it has recently been proved that spatially regular scalar fields which are non-minimally coupled 
to the Maxwell electromagnetic invariant can be supported by spinning and charged Kerr-Newman black holes. 
Motivated by this physically intriguing property of asymptotically flat black holes in composed Einstein-Maxwell-scalar field theories, 
we present a detailed {\it analytical} study of the physical and mathematical properties of the 
Maxwell electromagnetic invariant ${\cal F}_{\text{KN}}(r,\theta;M,a,Q)$ which characterizes the 
Kerr-Newman black-hole spacetime 
[here $\{r,\theta\}$ are respectively the radial and polar coordinates of 
the curved spacetime and $\{M,J=Ma,Q\}$ are respectively the mass, angular momentum, and 
electric charge parameters of the black hole]. 
It is proved that, for all Kerr-Newman black-hole spacetimes, the spin and charge 
dependent minimum value of the Maxwell electromagnetic invariant is attained on the equator of the black-hole surface. 
Interestingly, we reveal the physically important fact that Kerr-Newman spacetimes are characterized by two 
critical values of the dimensionless rotation parameter ${\hat a}\equiv a/r_+$ [here $r_+(M,a,Q)$ is the black-hole horizon radius], 
${\hat a}^{-}_{\text{crit}}=\sqrt{3-2\sqrt{2}}$ and ${\hat a}^{+}_{\text{crit}}=
\sqrt{5-2\sqrt{5}}$, which mark the boundaries between three qualitatively 
different spatial functional behaviors of the Maxwell electromagnetic invariant:
(i) Kerr-Newman black holes 
in the slow-rotation ${\hat a}<{\hat a}^{-}_{\text{crit}}$ regime are characterized by negative definite 
Maxwell electromagnetic invariants that increase monotonically towards spatial infinity, 
(ii) for black holes in the intermediate spin regime 
${\hat a}^{-}_{\text{crit}}\leq {\hat a}\leq{\hat a}^{+}_{\text{crit}}$, the positive global maximum of the Kerr-Newman 
Maxwell electromagnetic invariant is located at the black-hole poles, and 
(iii) Kerr-Newman black holes in the super-critical regime ${\hat a}>{\hat a}^{+}_{\text{crit}}$ 
are characterized by a non-monotonic spatial behavior of the Maxwell electromagnetic invariant 
${\cal F}_{\text{KN}}(r=r_+,\theta;M,a,Q)$ along the black-hole horizon with a spin and charge dependent global 
maximum whose polar angular location is characterized by the dimensionless functional relation ${\hat a}^2\cdot(\cos^2\theta)_{\text{max}}=5-2\sqrt{5}$.
\end{abstract}
\bigskip
\maketitle

\section{Introduction}

The most general black-hole solution of the non-linearly coupled Einstein-Maxwell 
field equations is described by the three-dimensional family \cite{Notetwo} of spinning and charged Kerr-Newman 
black holes. The Kerr-Newman spacetime is characterized by the curved line element \cite{ThWe,Chan,Notebl,Noteun}
\begin{eqnarray}\label{Eq1}
ds^2=-{{\Delta}\over{\rho^2}}(dt-a\sin^2\theta
d\phi)^2+{{\rho^2}\over{\Delta}}dr^2+\rho^2
d\theta^2+{{\sin^2\theta}\over{\rho^2}}\big[a
dt-(r^2+a^2)d\phi\big]^2\  ,
\end{eqnarray}
where the metric functions in (\ref{Eq1}) are given by the spatially-dependent functional expressions
\begin{equation}\label{Eq2}
\Delta\equiv r^2-2Mr+a^2+Q^2\ \ \ \ ; \ \ \ \ \rho^2\equiv r^2+a^2\cos^2\theta\  .
\end{equation}
The three physical parameters $\{M,J\equiv Ma,Q\}$ that characterize the Kerr-Newman spacetime (\ref{Eq1}) are respectively 
the black-hole mass, angular momentum, and electric charge \cite{Noteaq}.
The radial locations of the black-hole (outer and inner) horizons, 
\begin{equation}\label{Eq3}
r_{\pm}=M\pm\sqrt{M^2-a^2-Q^2}\  ,
\end{equation}
are determined by the roots of the metric function $\Delta(r)$. 

Interestingly, the electromagnetic tensor of the charged rotating Kerr-Newman 
black-hole spacetime (\ref{Eq1}) is characterized by a non-trivial spatially-dependent invariant  
\begin{equation}\label{Eq4}
{\cal F}\equiv F_{\mu\nu}F^{\mu\nu}\  ,
\end{equation}
known as the Maxwell electromagnetic invariant, whose spatial functional behavior is given 
by the expression \cite{MXI1,MXI2}
\begin{equation}\label{Eq5}
{\cal F}_{\text{KN}}(r,\theta;a,Q)=-{{2Q^2(r^4-6r^2a^2\cos^2\theta+a^4\cos^4\theta)}\over{(r^2+a^2\cos^2\theta)^4}}\  .
\end{equation}
For given values of the black-hole physical parameters $M$, $a$, and $Q$, the expression (\ref{Eq5}) for 
the Kerr-Newman Maxwell electromagnetic invariant 
is a two-dimensional function of the radial coordinate
\begin{equation}\label{Eq6}
r\in[M+\sqrt{M^2-a^2-Q^2},\infty]\
\end{equation}
and the polar angular coordinate
\begin{equation}\label{Eq7}
\theta\in[0,\pi]\ \Rightarrow \ \cos^2\theta\in[0,1]\  .
\end{equation}

The Maxwell electromagnetic invariant (\ref{Eq4}) of charged black-hole spacetimes 
has recently attracted the attention of physicists and mathematicians in the 
context of the physically intriguing phenomenon of black-hole spontaneous scalarization.  
In particular, it has been explicitly proved in \cite{Hersc1,Hersc2,Hodsc1,Hodsc2,Moh,Hodnrx} 
that the no-hair conjecture in black-hole physics \cite{NHC,JDB} can be 
violated in composed Einstein-Maxwell-scalar field theories whose action
\begin{equation}\label{Eq8}
S=\int
d^4x\sqrt{-g}\Big[R-2\nabla_{\alpha}\phi\nabla^{\alpha}\phi-f(\phi){\cal F}\Big]\
\end{equation}
contains a non-minimal direct coupling term $f(\phi){\cal F}$ 
between the scalar field and the Maxwell electromagnetic invariant (\ref{Eq4}) 
of the charged black-hole spacetime.

Intriguingly, it has been revealed \cite{Hersc1,Hersc2,Hodsc1,Hodsc2,Moh,Hodnrx} 
that the boundary between the familiar charged black-hole 
spacetimes of general relativity and scalarized (hairy) black-hole solutions 
of the composed Einstein-Maxwell-scalar field theory (\ref{Eq8}) is determined by the presence of a critical existence-curve of 
`cloudy' black-hole configurations, charged (Reissner-Nordstr\"om and Kerr-Newman) black holes that support 
linearized bound-state scalar field configurations which are non-minimally coupled to the 
Maxwell electromagnetic invariant (\ref{Eq4}) of the charged spacetime. 

Interestingly, the spatially-dependent Maxwell electromagnetic invariant (\ref{Eq4}) of charged black-hole spacetimes 
plays the role of an effective mass term in the generalized Klein-Gordon wave equation [see Eq. (\ref{Eq48}) below] of the 
Einstein-Maxwell-scalar field theory (\ref{Eq8}) \cite{Hersc1,Hersc2,Hodsc1,Hodsc2,Moh,Hodnrx}. 
In particular, depending on the values of the physical parameters $\{M,a,Q\}$ 
of the central supporting black hole, the spatially-dependent effective mass term may form 
a binding (attractive) black-hole-field potential well in the vicinity of the black-hole horizon, thus allowing 
the existence of bound-state cloudy scalar field configurations around the black hole. 

It is therefore of physical interest to explore the highly non-trivial two-dimensional functional properties of the Maxwell electromagnetic invariant 
${\cal F}_{\text{KN}}(r,\cos\theta;M,a,Q)$ that characterizes the spinning and charged 
Kerr-Newman black-hole spacetimes (\ref{Eq1}).

The main goal of the present paper is to study the 
physical and mathematical properties of the two-dimensional 
Maxwell electromagnetic invariant ${\cal F}_{\text{KN}}(r,\cos\theta;M,a,Q)$ in the exterior region (\ref{Eq6})
of spinning and charged Kerr-Newman black-hole spacetimes. 
In particular, we shall derive remarkably compact analytical formulas for the 
characteristic global extremum points of the Kerr-Newman Maxwell electromagnetic invariant (\ref{Eq5}). 
Intriguingly, using analytical techniques, we shall reveal the physically important fact that Kerr-Newman black holes 
are characterized by two charge-dependent critical rotation parameters, 
$\{(a/r_+)^{-}_{\text{crit}},(a/r_+)^{+}_{\text{crit}}\}$, that mark the boundaries 
between three qualitatively different spatial functional behaviors of the Kerr-Newman Maxwell electromagnetic invariant (\ref{Eq5}).
 
\section{Spatial behavior of the external two-dimensional Maxwell electromagnetic invariant 
${\cal F}_{\text{KN}}(r,\theta)$ within its domain of existence}

In the present section we shall explore the spatial properties of the two-dimensional 
Maxwell electromagnetic invariant (\ref{Eq5}) of the spinning and charged Kerr-Newman black-hole spacetime (\ref{Eq1}) 
in order to search for local extremum points and saddle points {\it within} the 
physically allowed external domain [see Eqs. (\ref{Eq6}) and (\ref{Eq7})]
\begin{equation}\label{Eq9}
r\in(M+\sqrt{M^2-a^2-Q^2},\infty)\ \ \ \ \text{with}\ \ \ \ \cos^2\theta\in(0,1)\  .
\end{equation}

The partial derivative
\begin{equation}\label{Eq10}
{{\partial {\cal F}_{\text{KN}}(r,\theta)}\over{\partial r}}=0\
\end{equation}
yields [for $(r,\cos^2\theta)\neq(0,0)$] the effectively quadratic equation
\begin{equation}\label{Eq11}
r^4-10r^2a^2\cos^2\theta+5a^4\cos^4\theta=0\  ,
\end{equation}
whereas the partial derivative
\begin{equation}\label{Eq12}
{{\partial {\cal F}_{\text{KN}}(r,\theta)}\over{\partial\theta}}=0\
\end{equation}
yields [for $(r,\cos^2\theta)\neq(0,0)$] the effectively quadratic equation \cite{Notecos0}
\begin{equation}\label{Eq13}
5r^4-10r^2a^2\cos^2\theta+a^4\cos^4\theta=0\  .
\end{equation}
The set of coupled equations (\ref{Eq11}) and (\ref{Eq13}) with $(r,\cos^2\theta)\neq(0,0)$ has 
no solutions. We therefore conclude 
that the two-dimensional Maxwell electromagnetic invariant (\ref{Eq5}) of the spinning and charged 
Kerr-Newman spacetime has no local extremum points or saddle points within 
the external black-hole domain (\ref{Eq9}).

\section{Functional behavior of the external Maxwell electromagnetic invariant ${\cal F}_{\text{KN}}(r,\theta)$ 
along its angular boundaries}

In the present section we shall analyze the functional behavior of the Kerr-Newman Maxwell electromagnetic invariant (\ref{Eq5}) 
along the angular boundaries [see Eq. (\ref{Eq7})] of the spinning and charged black-hole spacetime (\ref{Eq1}).  

\subsection{Analysis of the Maxwell electromagnetic invariant along the equatorial boundary $\cos^2\theta=0$}

Substituting the equatorial boundary relation 
\begin{equation}\label{Eq14}
\cos\theta=0\
\end{equation}
into the functional expression (\ref{Eq5}), one finds the remarkably compact behavior
\begin{equation}\label{Eq15}
{\cal F}_{\text{KN}}(r,\cos\theta=0)=-{{2Q^2}\over{r^4}}\  
\end{equation}
of the Maxwell electromagnetic invariant along the equatorial plane of the Kerr-Newman black hole. 
The expression (\ref{Eq15}) 
is a monotonically increasing function along the radial direction whose minimum (most negative) value is obtained at 
the outer horizon of the Kerr-Newman black hole:
\begin{equation}\label{Eq16}
{\cal F}_{\text{KN}}(r=M+\sqrt{M^2-a^2-Q^2},\cos\theta=0)=-{{2Q^2}\over{(M+\sqrt{M^2-a^2-Q^2})^4}}\  .
\end{equation}

\subsection{Analysis of the Maxwell electromagnetic invariant along the polar boundary $\cos^2\theta=1$}

Substituting the polar boundary relation 
\begin{equation}\label{Eq17}
\cos^2\theta=1\
\end{equation}
into the functional expression (\ref{Eq5}), one obtains the radially-dependent function
\begin{equation}\label{Eq18}
{\cal F}_{\text{KN}}(r,\cos^2\theta=1)=-{{2Q^2(r^4-6r^2a^2+a^4)}\over{(r^2+a^2)^4}}\ 
\end{equation}
for the Kerr-Newman Maxwell electromagnetic invariant. 
Intriguingly, we shall now prove that, depending on the values of the 
dimensionless physical parameters $\{a/M,Q/M\}$ that characterize the spinning and charged black hole, 
the polar electromagnetic expression (\ref{Eq18}) may have a non-trivial functional behavior 
along the radial direction. 

In particular, the radially-dependent Maxwell electromagnetic invariant (\ref{Eq18}) 
is characterized by two extremum points which are determined by the effectively quadratic equation
\begin{equation}\label{Eq19}
r^4-10r^2a^2+5a^4=0\  .
\end{equation}
From Eq. (\ref{Eq19}) one finds that the function (\ref{Eq18}), which describes the polar behavior of the 
Kerr-Newman Maxwell electromagnetic invariant, has one local radial minimum point,
\begin{equation}\label{Eq20}
r_{\text{min}}(a)=\sqrt{5+2\sqrt{5}}\cdot a\  ,
\end{equation}
that, in principle, can satisfy the requirement (\ref{Eq9}) \cite{Notex2s}. 

From Eqs. (\ref{Eq3}), (\ref{Eq6}), and (\ref{Eq20}) one deduces that, 
depending on the magnitude of the dimensionless spin parameter $a/r_+$ of the Kerr-Newman black-hole spacetime,  
%$a/M$ of a 
%Kerr-Newman black hole with a given non-zero value of its dimensionless electric charge parameter $Q/M$, 
the radially-dependent Maxwell electromagnetic invariant (\ref{Eq18}) has two qualitatively different functional behaviors:
\newline
{Case I:} From Eqs. (\ref{Eq3}), (\ref{Eq6}), and (\ref{Eq20}) one finds that, 
in the sub-critical regime \cite{Notecranmaxxx}
\begin{equation}\label{Eq21}
{{a}\over{r_+(M,a,Q)}}<\sqrt{1-2/\sqrt{5}}\
\end{equation}
of the Kerr-Newman black hole, which corresponds to the dimensionless charge-dependent spin regimes [see Eq. (\ref{Eq3})]
\begin{equation}\label{Eq22}
0<{{a}\over{M}}<{{\sqrt{5+2\sqrt{5}}+\sqrt{5+2\sqrt{5}-(6+2\sqrt{5})\cdot{(Q/M)}^2}}\over{6+2\sqrt{5}}}
\ \ \ \ \text{for}\ \ \ \ 0<{{Q}\over{M}}\leq\sqrt{{2}\over{\sqrt{5}}}\ 
\end{equation}
and
\begin{equation}\label{Eq23}
0<{{a}\over{M}}\leq\sqrt{1-\Big({{Q}\over{M}}\Big)^2}
\ \ \ \ \text{for}\ \ \ \ \sqrt{{2}\over{\sqrt{5}}}<{{Q}\over{M}}<1\  ,
\end{equation}
the minimum point (\ref{Eq20}) is located outside the radial region (\ref{Eq6}), 
in which case the polar expression (\ref{Eq18}) for the 
Kerr-Newman Maxwell electromagnetic invariant ${\cal F}_{\text{KN}}(r,\cos^2\theta=1)$ 
is a monotonically increasing function in the external radial region (\ref{Eq6}) of the black-hole 
spacetime (\ref{Eq1}) whose minimum value, 
\begin{eqnarray}\label{Eq24}
&{\cal F}_{\text{KN}}(r=M+\sqrt{M^2-a^2-Q^2},\cos^2\theta=1)=\nonumber \\ &
-{{2Q^2\big[(M+\sqrt{M^2-a^2-Q^2})^4-6(M+\sqrt{M^2-a^2-Q^2})^2a^2+a^4\big]}
\over{\big[(M+\sqrt{M^2-a^2-Q^2})^2+a^2\big]^4}}\  ,
\end{eqnarray}
is located on the outer horizon of the spinning and charged black hole. 
Note that the value of the Maxwell electromagnetic invariant is smaller at the minimum point (\ref{Eq16}) 
than at the minimum point (\ref{Eq24}). 
\newline
{Case II:} Taking cognizance of Eqs. (\ref{Eq3}), (\ref{Eq6}), and (\ref{Eq20}) one finds that, 
in the dimensionless super-critical regime 
\begin{equation}\label{Eq25}
{{a}\over{r_+(M,a,Q)}}\geq\sqrt{1-2/\sqrt{5}}\  ,
\end{equation}
which corresponds to the dimensionless spin regime [see Eq. (\ref{Eq3})] \cite{Notetwq}
\begin{equation}\label{Eq26}
{{\sqrt{5+2\sqrt{5}}+\sqrt{5+2\sqrt{5}-(6+2\sqrt{5})\cdot{(Q/M)}^2}}\over{6+2\sqrt{5}}}\leq
{{a}\over{M}}\leq \sqrt{1-{\Big({{Q}\over{M}}\Big)^2}}\ \ \ \ \text{for}\ \ \ \ 0<{{Q}\over{M}}\leq\sqrt{{2}\over{\sqrt{5}}}\
%\sqrt{{{5-\sqrt{5}}\over{32}}}+\sqrt{{{5-\sqrt{5}}\over{32}}-{{6-2\sqrt{5}}\over{16}}\cdot{\Big({{Q}\over{M}}\Big)^2}}\leq
%{{a}\over{M}}\leq \sqrt{1-{\Big({{Q}\over{M}}\Big)^2}}\
\end{equation}
of the Kerr-Newman spacetime, 
the Maxwell electromagnetic invariant ${\cal F}_{\text{KN}}(r,\cos^2\theta=1)$ has a local minimum point 
which is characterized by the radial relation (\ref{Eq20}) with the negative height
\begin{eqnarray}\label{Eq27}
{\cal F}_{\text{KN}}(r=r_{\text{min}},\cos^2\theta=1)=-{{5\sqrt{5}-11}\over{32a^4}}\cdot Q^2\  .
\end{eqnarray} 
One finds that, in the dimensionless regime (\ref{Eq25}) of the Kerr-Newman black-hole spin, 
the minimum point (\ref{Eq16}) is characterized by a Maxwell electromagnetic invariant whose value 
is smaller (more negative) than the corresponding value of the electromagnetic invariant at the minimum point (\ref{Eq27}). 

\section{Functional behavior of the Maxwell electromagnetic invariant ${\cal F}_{\text{KN}}(r,\theta)$ 
along its radial boundaries}

In the present section we shall explore the spatial properties of the Kerr-Newman Maxwell electromagnetic invariant (\ref{Eq5}) 
along the radial boundaries [see Eq. (\ref{Eq6})] of the external black-hole spacetime (\ref{Eq1}). 
We first point out that the 
Maxwell electromagnetic invariant of Kerr-Newman black-hole spacetimes is characterized 
by the trivial asymptotic behavior 
\begin{equation}\label{Eq28}
{\cal F}_{\text{KN}}(r\to\infty,\cos\theta)\to0^-\  .  
\end{equation}

Substituting the characteristic horizon boundary relation [see Eq. (\ref{Eq3})] 
\begin{equation}\label{Eq29}
r=r_+=M+\sqrt{M^2-a^2-Q^2}\  
\end{equation}
of the spinning and charged Kerr-Newman black holes into the functional expression (\ref{Eq5}), one finds the $\theta$-dependent 
relation
\begin{eqnarray}\label{Eq30}
&{\cal F}_{\text{KN}}(r=r_+,\cos\theta)= \nonumber \\ &
-{{2Q^2\big[(M+\sqrt{M^2-a^2-Q^2})^4-6(M+\sqrt{M^2-a^2-Q^2})^2a^2\cos^2\theta+a^4\cos^4\theta\big]}
\over{\big[(M+\sqrt{M^2-a^2-Q^2})^2+a^2\cos^2\theta\big]^4}}\ 
\end{eqnarray}
at the black-hole horizon. 

From the (rather cumbersome) expression (\ref{Eq30}) one deduces that 
the Maxwell electromagnetic invariant on the Kerr-Newman outer horizon 
may have a non-trivial (non-monotonic) functional dependence on the polar 
variable $\cos^2\theta$. 
In particular, using the composed dimensionless variable 
\begin{equation}\label{Eq31}
x\equiv {{a^2\cos^2\theta}\over{r^2_+(M,a,Q)}}\  ,
\end{equation}
one finds that the functional expression (\ref{Eq30}) can be written in the compact 
mathematical form
\begin{eqnarray}\label{Eq32}
{\cal F}_{\text{KN}}(r=r_+,\cos\theta)=
-{{2Q^2(1-6x+x^2)}\over{r^4_+(1+x)^4}}\  .
\end{eqnarray}
From Eq. (\ref{Eq32}) one finds the quadratic equation
\begin{equation}\label{Eq33}
x^2-10x+5=0\
\end{equation}
for the angular locations of the spin and charge dependent extremum points of the 
Maxwell electromagnetic invariant along the polar angular direction of the Kerr-Newman black-hole horizon. 
Interestingly, one finds that the maximum point 
\begin{equation}\label{Eq34}
x_{\text{max}}=5-2\sqrt{5}\
\end{equation}
of the Kerr-Newman Maxwell electromagnetic function (\ref{Eq32}) is the only physically acceptable 
solution of Eq. (\ref{Eq33}) that, in principle, can respect the angular condition (\ref{Eq7}) \cite{Notex2xx}. 

Taking cognizance of Eqs. (\ref{Eq3}), (\ref{Eq7}), (\ref{Eq31}), and (\ref{Eq34}) one deduces that, 
depending on the magnitude of the dimensionless spin parameter $a/r_+$ of the Kerr-Newman black-hole spacetime, 
%depending on the magnitude of the dimensionless spin parameter $a/M$ of a 
%Kerr-Newman black hole with a given non-zero value of its dimensionless electric charge parameter $Q/M$, 
the Maxwell electromagnetic invariant (\ref{Eq32}) along the black-hole horizon 
has two qualitatively different angular functional behaviors:

{Case I:} From Eqs. (\ref{Eq3}), (\ref{Eq7}), (\ref{Eq31}), and (\ref{Eq34}) one finds that, for spinning and charged 
Kerr-Newman black holes in the sub-critical regime \cite{Notecran}
\begin{equation}\label{Eq35}
{{a}\over{r_+(M,a,Q)}}<\sqrt{5-2\sqrt{5}}\  ,
\end{equation}
which corresponds to the dimensionless spin regimes [see Eq. (\ref{Eq3})]
%\begin{equation}\label{Eq28}
%{{a}\over{M}}<\Big({{a}\over{M}}\Big)_{\text{crit}}=
%{{\sqrt{5-2\sqrt{5}}\big[1+\sqrt{1-(6-2\sqrt{5})\cdot{(Q/M)}^2}\big]}\over{6-2\sqrt{5}}}\
%\end{equation}
\begin{equation}\label{Eq36}
0<{{a}\over{M}}<
\sqrt{{{5+\sqrt{5}}\over{32}}}+\sqrt{{{5+\sqrt{5}}\over{32}}-{{5-\sqrt{5}}\over{8}}\cdot{\Big({{Q}\over{M}}\Big)^2}}
\ \ \ \ \text{for}\ \ \ \ 0<{{Q}\over{M}}\leq\sqrt{2\sqrt{5}-4}\
\end{equation}
and
\begin{equation}\label{Eq37}
0<{{a}\over{M}}\leq\sqrt{1-\Big({{Q}\over{M}}\Big)^2}
\ \ \ \ \text{for}\ \ \ \ \sqrt{2\sqrt{5}-4}<{{Q}\over{M}}<1\
\end{equation}
of the rotating and charged Kerr-Newman black holes, 
the extremum point (\ref{Eq34}) is characterized by the non-physical relation $\cos^2\theta>1$ [see Eq. (\ref{Eq7})], 
in which case the Maxwell electromagnetic invariant (\ref{Eq32}) along the black-hole horizon 
is a monotonically increasing function of the variable $\cos^2\theta$ 
from the negative value
\begin{equation}\label{Eq38}
{\cal F}_{\text{KN}}(r=r_+,\cos\theta=0)=-{{2Q^2}\over{(M+\sqrt{M^2-a^2-Q^2})^4}}\ 
\end{equation}
to the value
\begin{eqnarray}\label{Eq39}
&{\cal F}_{\text{KN}}(r=r_+,\cos^2\theta=1)=\nonumber \\ &
-{{2Q^2\big[(M+\sqrt{M^2-a^2-Q^2})^4-6(M+\sqrt{M^2-a^2-Q^2})^2a^2+a^4\big]}
\over{\big[(M+\sqrt{M^2-a^2-Q^2})^2+a^2\big]^4}}\  .
\end{eqnarray}

It is important to point out that one learns from Eq. (\ref{Eq32}) that, in the regime 
\begin{equation}\label{Eq40}
x\geq3-2\sqrt{2}\  , 
\end{equation}
which, at the poles ($\cos^2\theta=1$) of the black-hole surface, corresponds to the dimensionless ratio [see Eq. (\ref{Eq31})]
\begin{equation}\label{Eq41}
{{a}\over{r_+(M,a,Q)}}\geq\sqrt{3-2\sqrt{2}}\  ,
\end{equation}
the Maxwell electromagnetic invariant (\ref{Eq39}) at the black-hole poles has a non-negative value. 
Thus, spinning and charged Kerr-Newman spacetimes in 
the dimensionless spin regime [see Eq. (\ref{Eq3})] \cite{Notetxc}
\begin{equation}\label{Eq42}
{{1+\sqrt{1-(4-2\sqrt{2})\cdot(Q/M)^2}}\over{2\sqrt{2}}}\leq{{a}\over{M}}\leq\sqrt{1-\Big({{Q}\over{M}}\Big)^2}
\ \ \ \ \text{for}\ \ \ \ 0<{{Q}\over{M}}\leq\sqrt{2\sqrt{2}-2}\  
\end{equation}
are characterized by Maxwell electromagnetic invariants whose values (\ref{Eq39}) at 
the poles of the black-hole surface are larger than the characteristic asymptotic value (\ref{Eq28}).

{Case II:} Interestingly, one finds that, for spinning and charged Kerr-Newman black holes 
in the super-critical regime 
\begin{equation}\label{Eq43}
{{a}\over{r_+(M,a,Q)}}\geq\sqrt{5-2\sqrt{5}}\  ,
\end{equation}
which corresponds to the dimensionless spin regime [see Eq. (\ref{Eq3})] \cite{Notetxcf}
\begin{equation}\label{Eq44}
\sqrt{{{5+\sqrt{5}}\over{32}}}+\sqrt{{{5+\sqrt{5}}\over{32}}-{{5-\sqrt{5}}\over{8}}\cdot{\Big({{Q}\over{M}}\Big)^2}}
\leq{{a}\over{M}}\leq\sqrt{1-\Big({{Q}\over{M}}\Big)^2}
\ \ \ \ \text{for}\ \ \ \ 0<{{Q}\over{M}}\leq\sqrt{2\sqrt{5}-4}\  ,
\end{equation}
the polar maximum point (\ref{Eq34}) of the function (\ref{Eq32}) is located 
within the physically allowed angular region (\ref{Eq7}), in which case the Maxwell electromagnetic invariant (\ref{Eq30}) 
is characterized by a non-trivial (non-monotonic) functional behavior along the polar angular direction 
of the Kerr-Newman black-hole horizon. 

In particular, taking cognizance of Eqs. (\ref{Eq31}) and (\ref{Eq34}), one finds 
the polar maximum point 
\begin{eqnarray}\label{Eq45}
(\cos^2\theta)_{\text{max}}=
\big({{r_+}\over{a}}\big)^2\cdot (5-2\sqrt{5})\
\end{eqnarray}
for the Maxwell electromagnetic invariant (\ref{Eq30}). 
The analytically derived relation (\ref{Eq45}) implies that, for a given value of the black-hole 
dimensionless charge parameter $Q/M$, the 
polar angle $\theta_{\text{max}}=\theta_{\text{max}}(a/M)$ (with $\theta_{\text{max}}\leq90^{\circ}$), 
which characterizes the maximum angular point of the Kerr-Newman 
Maxwell electromagnetic invariant (\ref{Eq30}), is 
a monotonically increasing function of the black-hole dimensionless spin parameter $a/M$.  

Taking cognizance of Eqs. (\ref{Eq32}), and (\ref{Eq34}), one obtains the functional expression
\begin{eqnarray}\label{Eq46}
{\cal F}_{\text{KN}}[r=r_+(a/M,Q/M),(\cos^2\theta)_{\text{max}}]=
{{Q^2}\over{r^4_+}}\cdot{{11+5\sqrt{5}}\over{32}}\  
\end{eqnarray}
for the maximal (most positive) value of the Maxwell electromagnetic invariant (\ref{Eq30}) that characterizes the spinning and 
charged Kerr-Newman black hole  along its horizon. 
It is of physical interest to stress the fact that, for a given value of the black-hole dimensionless charge parameter $Q/M$, 
the analytically derived functional relation (\ref{Eq46}) is 
a monotonically increasing function of the Kerr-Newman dimensionless rotation parameter $a/M$. 

%It is important to point out that one learns from Eq. (\ref{Eq}) that, in the regime 
%\begin{equation}\label{Eq45}
%x>3-2\sqrt{2}\  , 
%\end{equation}
%which, at the poles ($\cos^2\theta=1$) of the black-hole surface, corresponds to the dimensionless ratio [see Eq. (\ref{Eq})]
%\begin{equation}\label{Eq28}
%{{a}\over{r_+(M,a,Q)}}\geq\sqrt{3-2\sqrt{2}}\  ,
%\end{equation}
%the Maxwell electromagnetic invariant (\ref{Eq40}) at the black-hole poles has a non-negative value. 
%Thus, spinning and charged Kerr-Newman spacetimes in 
%the dimensionless spin regime [see Eq. (\ref{Eq})] \cite{Notetxc}
%\begin{equation}\label{Eq45}
%{{1+\sqrt{1-(4-2\sqrt{2})\cdot(Q/M)^2}}\over{2\sqrt{2}}}\leq{{a}\over{M}}\leq\sqrt{1-\Big({{Q}\over{M}}\Big)^2}
%\ \ \ \ \text{for}\ \ \ \ 0<{{Q}\over{M}}<\sqrt{2\sqrt{2}-2}\  
%\end{equation}
%are characterized by Maxwell electromagnetic invariants (\ref{Eq40}) whose values at 
%the poles of the black-hole surface are larger than the characteristic asymptotic value (\ref{Eq36}).

\section{Spinning and charged Kerr-Newman black holes supporting infinitesimally thin massive scalar rings}

In the present section we shall use the analytically derived results of the previous sections in order to reveal 
the physically interesting fact that Kerr-Newman black holes in 
Einstein-Maxwell-scalar field theories that contain a 
direct coupling between a massive scalar field and the Maxwell electromagnetic invariant (\ref{Eq5}) 
of the charged spacetime can support thin matter rings of the non-minimally coupled massive scalar fields.

The composed Einstein-Maxwell-massive-scalar field 
theory is characterized by the action \cite{Hersc1,Hersc2,Hodsc1,Hodsc2,Moh,Hodnrx} 
\begin{equation}\label{Eq47}
S=\int
d^4x\sqrt{-g}\Big[R-2\nabla_{\alpha}\phi\nabla^{\alpha}\phi-2\mu^2\phi^2-f(\phi){\cal F}\Big]\  ,
\end{equation}
where the physical parameter $\mu$ in the action (\ref{Eq47}) 
is the mass of the scalar field \cite{Notemuu}. 
The critical boundary between bald (scalarless) black-hole 
spacetimes and hairy black holes in the 
composed Einstein-Maxwell-scalar field theory (\ref{Eq47}) is determined by the presence of 
`cloudy' black-hole configurations \cite{Hersc1,Hersc2,Hodsc1,Hodsc2,Moh,Hodnrx}. 
These critical (marginally stable) configurations describe 
linearized bound-state non-minimally coupled scalar fields which are supported by the familiar 
(Reissner-Nordstr\"om and Kerr-Newman) charged black-hole spacetimes of general relativity. 

As we shall now demonstrate explicitly, the physically intriguing existence of composed 
black-hole-massive-scalar-field bound-state configurations in the Einstein-Maxwell-scalar field theory (\ref{Eq47}) 
is a direct outcome of the direct non-trivial interaction [see the coupling term $f(\phi){\cal F}$ in the action (\ref{Eq47})] 
between the massive scalar field $\phi$ and the spatially-dependent Maxwell 
electromagnetic invariant (\ref{Eq5}) of the charged spacetime.

The action (\ref{Eq47}), which characterizes the composed Einstein-Maxwell-nonminimally-coupled-scalar field theory, 
yields the Klein-Gordon equation \cite{Hersc1,Hersc2,Hodsc1,Hodsc2,Moh,Hodnrx}
\begin{equation}\label{Eq48}
\nabla^\nu\nabla_{\nu}\phi=\mu^2_{\text{eff}}\phi\
\end{equation}
for the scalar field, where the existence of the effective mass term 
\begin{equation}\label{Eq49}
\mu^2_{\text{eff}}(r,\theta;M,a,Q)=\mu^2-{1\over2}\alpha\cdot{\cal F}_{\text{KN}}(r,\theta;M,a,Q)\ 
\end{equation}
in the differential equation (\ref{Eq48}) is a direct outcome of the non-minimal coupling 
between the massive scalar field and the Maxwell electromagnetic invariant of the spacetime. 
The dimensionless physical parameter $\alpha<0$, which characterizes the weak-field 
functional behavior of the scalar coupling function \cite{Hersc1,Hersc2,Hodsc1,Hodsc2,Moh,Hodnrx},
\begin{equation}\label{Eq50}
f(\phi)=1-\alpha\phi^2\  ,
\end{equation}
determines the strength of the direct scalar-field-Maxwell-electromagnetic-invariant interaction.  

Intriguingly, from Eq. (\ref{Eq49}) one learns that, depending on the relative magnitudes of the physical 
parameters $\alpha$ and $\mu$ that characterize the composed Einstein-Maxwell-nonminimally-coupled-massive-scalar field 
theory (\ref{Eq47}), the effective mass term (\ref{Eq49}) in the Klein-Gordon scalar differential equation (\ref{Eq48}) 
may become negative (attractive) in the vicinity of the outer horizon (\ref{Eq3}) 
of the spinning and charged Kerr-Newman black hole. 

The presence of an effective negative mass term (an effective potential well) 
in the scalar Klein-Gordon equation (\ref{Eq48}) provides a necessary condition for the existence of 
scalar clouds (linearized bound-state configurations of the non-minimally coupled scalar fields) 
in the spinning and charged Kerr-Newman black-hole spacetime (\ref{Eq1}) \cite{Donnw1,ChunHer,Hodca,Herkn}. 
%\cite{Dons,Donnw1,Donnw2,ChunHer,Hodca,Herkn}. 
In particular, in the dimensionless large-mass regime 
[or equivalently, in the large-coupling $|\alpha|\gg1$ regime, see Eq. (\ref{Eq53}) below]
\begin{equation}\label{Eq51}
\mu r_+\gg1\
\end{equation}
of the non-minimally coupled scalar field, 
the {\it onset} of the black-hole spontaneous scalarization phenomenon 
is determined by the critical functional relation \cite{Hodca,Herkn,Hodjp}
\begin{equation}\label{Eq52}
\text{min}\{\mu^2_{\text{eff}}(r,\theta;M,a,Q)\}\to 0^{-}\  ,
\end{equation}
which characterizes the parameter-dependent boundary between repulsive 
and attractive effective black-hole-scalar-field interaction potentials. 

From Eqs. (\ref{Eq16}), (\ref{Eq49}), and (\ref{Eq52}) one deduces that 
the onset of the spontaneous scalarization phenomenon in the spinning and charged Kerr-Newman black-hole 
spacetime (\ref{Eq1}) is determined by the critical functional relation 
\begin{equation}\label{Eq53}
-{{Q^2}\over{(M+\sqrt{M-a^2-Q^2})^4}}\cdot{{\alpha}\over{\mu^2}}\to 1^{+}\  .
\end{equation}
In particular, the effective mass term (\ref{Eq49}), 
which characterizes the composed Kerr-Newman-black-hole-nonminimally-coupled-massive-scalar-field system,  
becomes {\it negative} (and therefore represents an attractive near-horizon 
black-hole-scalar-field interaction) in the narrow equatorial 
{\it ring} which is characterized by the relations $r\to r_+(M,a,Q)$ with $\theta\to\pi/2$. 
This physically intriguing observation implies that spinning and charged Kerr-Newman black-hole spacetimes 
can support infinitesimally thin equatorial clouds of the massive scalar fields (massive scalar rings) which 
are characterized, in the dimensionless large-mass regime (\ref{Eq51}) of the 
composed black-hole-scalar-field system, by the critical ratio (\ref{Eq53}). 

\section{Summary and discussion}

Motivated by the recent intriguing discovery of cloudy black holes \cite{Notekb} in 
composed Einstein-Maxwell-nonminimally-coupled-scalar field theories that support bound-state scalar matter 
configurations with a direct coupling to the Maxwell electromagnetic invariant of the charged spacetime \cite{Hersc1,Hersc2,Hodsc1,Hodsc2,Moh,Hodnrx}, 
we have explored the physical and mathematical properties of the spatially-dependent 
Maxwell electromagnetic invariant ${\cal F}_{\text{KN}}(r,\cos\theta;M,a,Q)$ [see Eq. (\ref{Eq5})] of the spinning and charged 
Kerr-Newman black-hole spacetime. 

The main {\it analytical} results derived in this paper and their physical implications are as follows:

(1) It has been proved that the global minimum point \cite{Notering} of the Kerr-Newman Maxwell electromagnetic 
invariant in the exterior region $r\in[r_+(M,a,Q),\infty]$ of the black-hole spacetime is always 
(that is, for all physically allowed values of the dimensionless physical parameters $a/M$ and $Q/M$) 
located on the black-hole horizon. 
In particular, from Eqs. (\ref{Eq16}), (\ref{Eq24}), and (\ref{Eq27}) 
one deduces that the global minimum (most negative) value 
\begin{equation}\label{Eq54}
\text{min}\Big\{M^2\cdot{\cal F}_{\text{KN}}(r\in[r_+,\infty],\cos^2\theta;{\bar a},{\bar Q})\Big\}=
-{{2{\bar Q}^2}\over{(1+\sqrt{1-{\bar a}^2-{\bar Q}^2})^4}}\
\end{equation}
of the Maxwell electromagnetic invariant is 
located on the equator ($\theta_{\text{max}}=90^{\circ}$) of the Kerr-Newman black-hole surface, 
where 
\begin{equation}\label{Eq55}
{\bar a}\equiv {{a}\over{M}}\ \ \ \ ; \ \ \ \ {\bar Q}\equiv{{Q}\over{M}}\  .
\end{equation}

(2) From the analytically derived functional expression (\ref{Eq54}) one deduces that, 
for a given value ${\bar Q}$ of the black-hole electric charge,  the global minimum value 
of the Kerr-Newman Maxwell electromagnetic invariant is a monotonically decreasing function 
of the black-hole spin parameter ${\bar a}$. 
In particular, from Eq. (\ref{Eq54}) one finds the dimensionless value  
\begin{equation}\label{Eq56}
\text{min}\Big\{M^2\cdot{\cal F}_{\text{KN}}(r\in[r_+,\infty],\cos^2\theta;{\bar a}^2+{\bar Q}^2=1)\Big\}=
-2{\bar Q}^2\
\end{equation}
for extremal Kerr-Newman black holes. 
This is the smallest (that is, the most negative) value of the Maxwell electromagnetic invariant 
that characterizes Kerr-Newman black holes 
with a given value of the dimensionless electric charge parameter ${\bar Q}$. 

(3) We have revealed the physically intriguing fact that the location of the global maximum point, which characterizes the spatial behavior 
of the Maxwell electromagnetic invariant of the spinning and charged Kerr-Newman black-hole spacetime, has a highly 
non-trivial functional dependence on the dimensionless physical parameter
\begin{equation}\label{Eq57}
{\hat a}\equiv {{a}\over{r_+}}\  .
\end{equation} 
In particular, we have revealed the existence of two critical values [see Eqs. (\ref{Eq41}) and (\ref{Eq43})],
\begin{equation}\label{Eq58}
{\hat a}^{-}_{\text{crit}}=\sqrt{3-2\sqrt{2}}\
\end{equation}
and
\begin{equation}\label{Eq59}
{\hat a}^{+}_{\text{crit}}=\sqrt{5-2\sqrt{5}}\  ,
\end{equation}
for the dimensionless rotation parameter (\ref{Eq57}) of the Kerr-Newman black-hole spacetime. 

The critical rotation parameters (\ref{Eq58}) and (\ref{Eq59}) 
mark the boundaries between three qualitatively different spatial functional behaviors of the Kerr-Newman 
Maxwell electromagnetic invariant: 
\newline
(i) Kerr-Newman black holes in the dimensionless sub-critical spin 
regime ${\hat a}<{\hat a}^{-}_{\text{crit}}$ are characterized by negative definite Maxwell electromagnetic 
invariants that attain their global maxima asymptotically at spatial infinity [see Eq. (\ref{Eq28})]. 
\newline
(ii) Spinning and charged Kerr-Newman black holes in the intermediate spin 
regime ${\hat a}^{-}_{\text{crit}}\leq {\hat a}\leq{\hat a}^{+}_{\text{crit}}$ 
are characterized by Maxwell electromagnetic invariants whose 
global positive maxima are located at the two poles of the black-hole surface:
\newline
\begin{eqnarray}\label{Eq60}
(\cos^2\theta)_{\text{max}}=1\ \ \ \ \ \text{for}\ \ \ \ \ {\hat a}^{-}_{\text{crit}}\leq {\hat a}\leq{\hat a}^{+}_{\text{crit}}\  .
\end{eqnarray}
\newline
(iii) Kerr-Newman black holes in the super-critical spin regime
${\hat a}>{\hat a}^{+}_{\text{crit}}$ have spatially-dependent 
Maxwell electromagnetic invariants which are characterized by non-monotonic functional behaviors along 
the polar angular direction of the black-hole surface. 
In particular, the global maxima that characterize the spin and charge dependent Maxwell electromagnetic invariants 
of these super-critical Kerr-Newman black holes are determined by the analytically derived 
compact scaling relation [see Eq. (\ref{Eq45})]
\begin{eqnarray}\label{Eq61}
{\hat a}^2\cdot(\cos^2\theta)_{\text{max}}=5-2\sqrt{5}
\ \ \ \ \text{for}\ \ \ \ {\hat a}\geq{\hat a}^{+}_{\text{crit}}\  . 
\end{eqnarray}

(4) The analytically derived expression (\ref{Eq61}) reveals the fact that
%,for a given value of the black-hole electric charge parameter ${\bar Q}$, 
the polar angle $\theta_{\text{max}}$ (with $\theta_{\text{max}}\leq90^{\circ}$), 
which characterizes the maximum angular point of the Kerr-Newman Maxwell 
electromagnetic invariant in the charge-dependent super-critical regime ${\hat a}\geq{\hat a}^{+}_{\text{crit}}$, is 
a monotonically increasing function of the black-hole dimensionless spin parameter ${\hat a}$. 
In particular, from Eq. (\ref{Eq61}) one finds the asymptotic behavior \cite{Noteth2}
\begin{equation}\label{Eq62}
\theta^{-}_{\text{max}}({\hat a}\to1^{-})\simeq43.4^{\circ}\ 
\end{equation}
in the limit of maximally-spinning Kerr-Newman black holes with an infinitesimally small non-zero electric charge. 
Interestingly, the value $\theta^{-}_{\text{max}}\simeq43.4^{\circ}$ is the {\it largest} polar angle (with $\theta_{\text{max}}\leq90^{\circ}$) that characterizes the global maximum points of the Maxwell 
electromagnetic invariants of spinning and charged curved Kerr-Newman black-hole spacetimes.  

(5) From the analytically derived functional expressions (\ref{Eq28}), (\ref{Eq39}), and (\ref{Eq46}) 
one finds that, in the exterior region (\ref{Eq6}) of the spinning and charged Kerr-Newman black-hole spacetime (\ref{Eq1}), 
the spatially-dependent Maxwell electromagnetic invariant is characterized by the global dimensionless maximum
\begin{eqnarray}\label{Eq63}
&\text{max}\Big\{M^2\cdot{\cal F}_{\text{KN}}(r\in[r_+,\infty],\cos^2\theta;{\bar a},{\bar Q})\Big\}=
\nonumber\\ 
&
\begin{cases}
0^{-} & \ \ \ \text{for}\ \ \ \ \ \ {\hat a}<{\hat a}^{-}_{\text{crit}}\ \\
-{{2{\bar Q}^2\big[(1+\sqrt{1-{\bar a}^2-{\bar Q}^2})^4-6(1+\sqrt{1-{\bar a}^2-{\bar Q}^2})^2{\bar a}^2+{\bar a}^4\big]}
\over{\big[(1+\sqrt{1-{\bar a}^2-{\bar Q}^2})^2+{\bar a}^2\big]^4}} & \ \ \ \text{for}\ \ \ \ \ 
\ {\hat a}^{-}_{\text{crit}}\leq {\hat a}\leq{\hat a}^{+}_{\text{crit}}\ \\
{{{\bar Q}^2}\over{(1+\sqrt{1-{\bar a}^2-{\bar Q}^2})^4}}\cdot{{11+5\sqrt{5}}\over{32}} & 
\ \ \ \text{for}\ \ \ \ \ \ {\hat a}\geq{\hat a}^{+}_{\text{crit}}\  . 
\end{cases}
\end{eqnarray}

(6) Interestingly, one deduces from the functional relation (\ref{Eq63}) that, 
for a given value of the black-hole electric charge parameter ${\bar Q}$, the global maximum value 
of the Kerr-Newman Maxwell electromagnetic invariant in the super-critical spin 
regime ${\hat a}\geq{\hat a}^{+}_{\text{crit}}$ is a monotonically increasing function of the black-hole 
spin parameter ${\bar a}$. 
In particular, from Eq. (\ref{Eq63}) one finds the dimensionless functional expression  
\begin{equation}\label{Eq64}
\text{max}\Big\{M^2\cdot{\cal F}_{\text{KN}}(r\in[r_+,\infty],\cos^2\theta;{\bar a}^2+{\bar Q}^2=1)\Big\}=
{\bar Q}^2\cdot{{11+5\sqrt{5}}\over{32}}\
\end{equation}
for extremal spinning and charged Kerr-Newman black holes that belong to the 
super-critical ${\hat a}\geq{\hat a}^{+}_{\text{crit}}$ regime. 
This is the largest value of the Maxwell electromagnetic invariant that characterizes Kerr-Newman black-hole spacetimes 
with a given value of the dimensionless electric charge parameter ${\bar Q}$. 

(7) From Eq. (\ref{Eq64}) one learns that the Maxwell electromagnetic invariant 
in the super-critical regime ${\hat a}\geq{\hat a}^{+}_{\text{crit}}$ is a monotonically increasing function of the 
black-hole dimensionless charge parameter ${\bar Q}$. 
Taking cognizance of Eq. (\ref{Eq44}), one finds that the largest allowed value of the black-hole electric charge parameter 
in the super-critical regime ${\hat a}\geq{\hat a}^{+}_{\text{crit}}$ is given by the dimensionless expression
\begin{equation}\label{Eq65}
{\bar Q}_{\text{max}}=\sqrt{2\sqrt{5}-4}\ \ \ \ \text{for}\ \ \ \ {\hat a}\geq{\hat a}^{+}_{\text{crit}}\  .
\end{equation}
We therefore conclude that, among all Kerr-Newman black-hole spacetimes, 
the largest possible value of the external Maxwell electromagnetic invariant is 
obtained for the extremal spinning and charged black hole with the physical parameters 
\begin{equation}\label{Eq66}
({\bar a},{\bar Q})=\Big(\sqrt{5-2\sqrt{5}},\sqrt{2\sqrt{5}-4}\Big)\  .
\end{equation}
This unique black-hole spacetime is characterized by the remarkably compact analytically derived expression
\begin{equation}\label{Eq67}
\text{max}\Big\{M^2\cdot{\cal F}_{\text{KN}}(r\in[r_+,\infty],\cos^2\theta)\Big\}=
{{3+\sqrt{5}}\over{16}}\  
\end{equation}
for the highest (most positive) value of the spatially-dependent Maxwell electromagnetic invariant. 

(8) Finally, we have revealed the physically interesting fact that, in the dimensionless regime (\ref{Eq51}) of large field masses, 
the spinning and charged Kerr-Newman black holes can support infinitesimally thin equatorial matter configurations which are 
made of massive scalar fields with a non-minimal direct coupling to the Maxwell electromagnetic 
invariant of the charged spacetime. The supported massive scalar rings (scalar clouds) of the composed 
Einstein-Maxwell-nonminimally-coupled-massive-scalar field theory (\ref{Eq47}) 
are located on the equator of the Kerr-Newman black-hole surface and 
are characterized by the remarkably compact critical functional relation [see Eqs. (\ref{Eq3}) 
and (\ref{Eq53})] 
\begin{equation}\label{Eq68}
-{{Q^2}\over{r^4_+}}\cdot{{\alpha}\over{\mu^2}}\to 1^{+}\  .
\end{equation}\  .

\bigskip
\noindent
{\bf ACKNOWLEDGMENTS}
\bigskip

This research is supported by the Carmel Science Foundation. I would
like to thank Yael Oren, Arbel M. Ongo, Ayelet B. Lata, and Alona B.
Tea for helpful discussions.

%\newpage


\begin{thebibliography}{99}

\bibitem{Notetwo} We use here the term `three-dimensional' in order 
to emphasize the fact that stationary Kerr-Newman black holes are characterized by three asymptotically measured conserved 
physical parameters: the  
mass $M$, the angular momentum $J\equiv Ma$, and the electric charge $Q$ of the black-hole spacetime.  

\bibitem{ThWe} C. W. Misner, K. S. Thorne, and J. A. Wheeler, {\it Gravitation}, (W. H. Freeman, San Francisco, 1973).

\bibitem{Chan} S. Chandrasekhar, {\it The Mathematical Theory of Black
Holes}, (Oxford University Press, New York, 1983).

\bibitem{Notebl} The curved line element (\ref{Eq1}) is expressed in terms of the 
Boyer-Lindquist spacetime coordinates $(t,r,\theta,\phi)$.

\bibitem{Noteun} We use natural units in which $8\pi G=c=\hbar=1$.

\bibitem{Noteaq} We shall assume, without loss of generality, the relations $a\geq0$ and $Q\geq0$ for the 
physical parameters of the spinning and charged Kerr-Newman black-hole spacetime. 

\bibitem{MXI1} T. Adamo and E.T. Newman, The Kerr-Newman metric: A Review, arXiv:1410.6626 .

\bibitem{MXI2} I. Dymnikova and E. Galaktionov, 
Advan. in Math. Phys. {\bf 2017}, 1035381 (2017) [https://www.hindawi.com/journals/amp/2017/1035381/]; 
I. Dymnikova and E. Galaktionov, Universe {\bf 5}, 205 (2019).

\bibitem{Hersc1} C. A. R. Herdeiro, E. Radu, N. Sanchis-Gual, and J. A. Font, Phys. Rev. Lett. {\bf 121}, 101102 (2018).

\bibitem{Hersc2} P. G. S. Fernandes, C. A. R. Herdeiro, A. M. Pombo, E. Radu, and N. Sanchis-Gual,
Class. Quant. Grav. {\bf 36}, 134002 (2019) [arXiv:1902.05079].

\bibitem{Hodsc1} S. Hod, Phys. Lett. B {\bf 798}, 135025 (2019) [arXiv:2002.01948].

\bibitem{Hodsc2} S. Hod, Phys. Rev. D {\bf 101}, 104025 (2020) [arXiv:2005.10268].

\bibitem{Moh} M. Khodadi, A. Allahyari, S. Vagnozzi, and D. F. Mota, JCAP {\bf 09}, 026 (2020) [arXiv:2005.05992].

\bibitem{Hodnrx} S. Hod, The Euro. Phys. Jour. C {\bf 80}, 1150 (2020).

\bibitem{NHC} R. Ruffini and J. A. Wheeler, Physics Today {\bf 24}, 30 (1971).

\bibitem{JDB} J. D. Bekenstein, Contribution to the 2nd International Sakharov Conference on Physics, 216-219 
[e-Print: gr-qc/9605059].

\bibitem{Notecos0} It is important to emphasize that 
in the present section we analyze the spatial functional behavior of the Kerr-Newman Maxwell electromagnetic invariant 
inside its domain of existence, in which case $\cos\theta\neq0$ [see Eq. (\ref{Eq9})].

\bibitem{Notex2s} Note that the second solution of the effectively quadratic equation (\ref{Eq19}), 
$r_{\text{max}}(a)=\sqrt{5-2\sqrt{5}}\cdot a$, 
which corresponds to a local maximum radial point of the Maxwell electromagnetic function (\ref{Eq18}), is 
characterized by the relation $r_{\text{max}}(a)<a$. 
Thus, it cannot respect the characteristic radial relation (\ref{Eq9}) of the external Kerr-Newman spacetime [see Eq. (\ref{Eq3})]. 

\bibitem{Notecranmaxxx} Note that the dimensionless critical ratio (\ref{Eq21}) is obtained by substituting 
the marginally allowed Kerr-Newman value $r=r_+=M+\sqrt{M^2-a^2-Q^2}$ of the external radial coordinate 
into the analytically derived relation (\ref{Eq20}). 

\bibitem{Notetwq} Note that the two inequalities in (\ref{Eq26}) can be satisfied simultaneously in 
the dimensionless charge regime $0<Q/M\leq\sqrt{{{5+\sqrt{5}}\over{8}}}$ of the 
Kerr-Newman black hole. In particular, the two inequalities in (\ref{Eq26}) can be satisfied simultaneously in 
the entire regime $0<{{Q}/{M}}\leq\sqrt{{2}\over{\sqrt{5}}}$. 
%\sqrt{{{4+2\sqrt{5}}\over{5+2\sqrt{5}}}}$. 
%\sqrt{{{2}\over{\sqrt{5}}}}$

\bibitem{Notex2xx} Note that the second solution of the quadratic equation (\ref{Eq33}), $x_{\text{min}}=5+2\sqrt{5}$, 
which correspond to a local minimum angular point of the Kerr-Newman Maxwell electromagnetic function (\ref{Eq32}), 
is larger than $1$. 
Thus, it cannot respect the angular relation (\ref{Eq7}) [see Eq. (\ref{Eq31})].

\bibitem{Notecran} Note that the dimensionless critical ratio (\ref{Eq35}) is obtained by substituting 
the marginally allowed angular value $\cos^2\theta=1$ 
into the expression (\ref{Eq34}) for the polar location of the maximum angular 
point of the Kerr-Newman Maxwell electromagnetic invariant (\ref{Eq32}) along the black-hole horizon. 

\bibitem{Notetxc} Note that the two inequalities in (\ref{Eq42}) can be satisfied simultaneously in 
the dimensionless charge regime $0<Q/M\leq\sqrt{{{2+\sqrt{2}}\over{4}}}$ of the 
Kerr-Newman black hole. In particular, the two inequalities in (\ref{Eq42}) can be satisfied simultaneously in 
the entire regime $0<{{Q}/{M}}<\sqrt{2\sqrt{2}-2}$. 

\bibitem{Notetxcf} Note that the two inequalities in (\ref{Eq44}) can be satisfied simultaneously in 
the dimensionless charge regime $0<Q/M\leq\sqrt{{{1+\sqrt{5}}\over{4}}}$ of the 
Kerr-Newman black hole. In particular, the two inequalities in (\ref{Eq44}) can be satisfied simultaneously in 
the entire regime $0<{{Q}\over{M}}<\sqrt{2\sqrt{5}-4}$.

\bibitem{Notemuu} Note that the field parameter $\mu$ stands for $\mu/\hbar$. Hence, this 
physical parameter has the dimensions of $\text{length}^{-1}$.

%\bibitem{Donnw2} D. D. Doneva, L. G. Collodel, C. J. Krüger, S. S. Yazadjiev, 
%The Eur. Phys. Jour. C {\bf 80}, 1205 (2020) [arXiv:2009.03774]. 

\bibitem{Donnw1} D. D. Doneva, L. G. Collodel, C. J. Krüger, S. S. Yazadjiev, 
Phys. Rev. D {\bf 102}, 104027 (2020) [arXiv:2008.07391].

%\bibitem{Dons} D. D. Doneva and S. S. Yazadjiev, Phys. Rev. Lett. {\bf 120}, 131103 (2018).

\bibitem{ChunHer} P. V. P. Cunha, C. A. R. Herdeiro, and E. Radu, Phys. Rev. Lett. {\bf 123}, 011101 (2019).

\bibitem{Hodca} S. Hod, Phys. Rev. D {\bf 102}, 084060 (2020) [arXiv:2006.09399].

\bibitem{Herkn} C. A. R. Herdeiro, A. M. Pombo, and E. Radu, Universe {\bf 7}, 483 (2021) [arXiv:2111.06442].

\bibitem{Hodjp} S. Hod, Jour. of High Energy Phys. {\bf 02}, 039 (2022) [arXiv:2201.03503].

\bibitem{Notekb} It is important to emphasize again that we use here the 
term `cloudy' black-hole spacetimes in order to describe 
black holes that support spatially regular linearized scalar fields. The physical significance 
of these composed black-hole-field cloudy configurations stems from the fact that, in the Einstein-Maxwell-scalar field theory (\ref{Eq47}), 
they mark the critical boundary between bald (scalarless) Kerr-Newman black-hole spacetimes and composed non-linearly 
coupled black-hole-scalar-field hairy configurations.  

\bibitem{Notering} Note that, due to the azimuthal symmetry that characterizes the spinning and charged 
Kerr-Newman black-hole spacetime (\ref{Eq1}), 
the spatially-dependent Maxwell electromagnetic invariant (\ref{Eq5}) is characterized by extremum rings with $\phi\in[0,2\pi]$. 

\bibitem{Noteth2} Note that, due to the reflection symmetry that characterizes the spinning and charged Kerr-Newman 
black-hole spacetime (\ref{Eq1}), the second angular maximum point (with $\theta_{\text{max}}>90^{\circ}$) 
of the spatially-dependent 
Maxwell electromagnetic invariant (\ref{Eq5}) is given by the 
simple functional relation $\theta^{+}_{\text{max}}=180^{\circ}-\theta^{-}_{\text{max}}$. 

\end{thebibliography}
\end{document}